\documentclass[11pt]{article}

\usepackage[T1]{fontenc}
\usepackage[utf8]{inputenc}
\usepackage{amsmath,amssymb}
\usepackage{graphicx}
\usepackage{url}
\usepackage{hyperref}
\usepackage{lineno}
\usepackage{microtype}
\usepackage{subcaption}
\usepackage{wrapfig}
\usepackage{gensymb}
\usepackage{array,multicol}
\usepackage{units}
\usepackage{setspace}
\usepackage{soul}
\usepackage{orcidlink}

\usepackage[letterpaper,left=0.75in,right=0.75in,top=1in,bottom=1in]{geometry}

\usepackage{float}

\nolinenumbers

\title{Mechanics-guided parametric modeling of intranasal spray devices and formulations for targeted drug delivery to the nasopharynx}
\author{
Md Tariqul Hossain\footnote{~South Dakota State University, Department of Mechanical Engineering, Brookings, South Dakota, United States} \and
Abir Malakar\footnotemark[1] \and
Mohammad Yeasin\footnotemark[1] \and
William O'Connell\footnotemark[1] \and
Mohammad Mehedi Hasan Akash\footnotemark[1]~\footnote{~Florida State University FAMU-FSU College of Engineering, Department of Mechanical Engineering, Tallahassee, Florida, United States} \and
Azadeh Borojeni\footnotemark[1] \and
Devranjan Samanta\footnote{~Indian Institute of Technology Ropar, Department of Mechanical Engineering, Rupnagar, Punjab, India} \and
Gerallt Williams\footnote{~Aptar Pharma, Route des Falaise, Le Vaudreuil, France} \and
Joshua Reineke\footnote{~South Dakota State University, Department of Pharmaceutical Sciences, Brookings, South Dakota, Umited States} \and
Gonçalo Farias\footnotemark[4] \and
Sunghwan Jung\footnote{~Cornell University, Department of Biological and Environmental Engineering, Ithaca, New York, United States} \and
Julie Suman\footnote{~Aptar Pharma, Congers, New York, USA} \and
Saikat Basu\footnotemark[1]~~\footnote{~Corresponding author: Saikat.Basu@sdstate.edu (\href{https://sites.google.com/view/basulab-tour/home}{Biomedical and Bioinspired Fluid Dynamics Lab})}~~~\orcidlink{0000-0003-1464-8425}
}

\date{}
\begin{document}
\maketitle

\begin{abstract}
\noindent Improving the efficacy of nasal sprays by enhancing targeted drug delivery to intra-airway tissue sites prone to infection onset is hypothesized to be achievable through an optimization of key device and formulation parameters, such as the sprayed droplet sizes, the spray cone angle, and the formulation density. This study focuses on the nasopharynx, a primary locus of early viral entry, as the optimal target for intranasal drug delivery. Three-dimensional anatomical upper airway geometries reconstructed from high-resolution computed tomography scans were used to numerically evaluate a cone injection approach, with inert particles mimicking the motion of sprayed droplets within an underlying inhaled airflow field. We have considered monodisperse sprayed particles sized between 10 – 50 $\mu$m, six densities ranging from  1.0 – 1.5 g/ml for the constituent formulation, and twelve plume angles spanning 15$\degree$ – 70$\degree$ subtended by the spray jet at the nozzle position. Large Eddy Simulation-based modeling of the inhaled airflow physics within the anatomical domains was coupled with a Lagrangian particle-tracking framework to derive the drug deposition trend at the nasopharynx. The resulting three-dimensional deposition contour map, obtained by interpolating the outcomes for the discrete test parameters,  revealed that nasopharyngeal deposition peaked for droplet sizes 25 – 45 $\mu$m and plume angles $\leq$ 30$\degree$, when the nasopharyngeal deposition rates are averaged over the test airway geometries and formulation densities. In addition, the formulation density of 1.0 g/ml yielded the highest mean deposition rate, over the tested range of sprayed particle sizes and plume angles. The findings were experimentally validated through representative physical spray tests conducted in a 3D-printed replica of one of the test geometries and collectively demonstrate that rational optimization of the intranasal spray design is attainable, with substantial enhancement of targeted drug delivery to the nasopharynx.\\

\end{abstract}


\section{Introduction}

Respiratory viral infections, including influenza, COVID-19, and the common cold, continue to pose major global public health challenges \cite{volpe2023viral}. Effective treatment during the initial phase of infection and, in general, prevention are crucial to reducing the impact of these diseases. Intranasal drug delivery systems, especially intranasal sprays \cite{pires2009csps,popper2023lar,wu2025brain}, have emerged as a promising method for delivering targeted therapeutic agents, vaccines, and antiviral medications directly to the infected tissue sites along the airway \cite{afkhami2022respiratory,mi2024ecm,jin2024cti,banella2025ddtr}.

\begin{figure}
\centering
\includegraphics[width=0.5\textwidth]{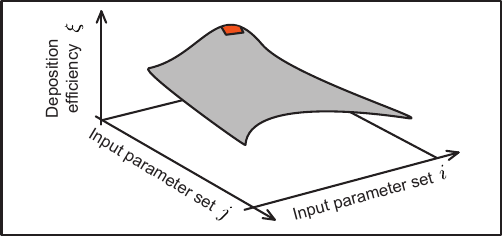}
\caption{{\bf Envisioned 3D contour map:} The vertical axis represents the deposition efficiency ($\xi$) and the two horizontal axes represent two controllable input delivery parameters (e.g., spray plume angle of the delivery device, particle sizes of drug solution, nozzle exit speed of the delivery device, viscosity of the solution, delivery axis orientation, drug solution density etc.). For this study, the two chosen input variable axes are particle size range ($d$; 10 -- 50~$\mu$m) and a range of plume angle of delivery device ($\theta$; 15\degree -- 70\degree), with 6 different drug solution densities (1.0 -- 1.5 g/ml). 24 different contour plots were generated for each test airway (number of airways is 4; 2 for each subject). Hence, the results comprise 24 generated plots (see Figs.~\ref{fig6} and \ref{fig7}). By averaging the data across geometries and test formulation densities (see Fig.~\ref{fig8} and specifically Fig.~\ref{fig9}), this study identifies the equivalent of the red region (illustrated on the cartoon parametric map above) that captures the suitable conditions for maximal targeted deposition efficiency.}
\label{fig1}
\end{figure}

The nasopharynx—the upper part of the pharynx located at the back of the nose—serves as a critical hotspot for initial respiratory infections via inhaled transmission \cite{matheson2020science,hou2020cell,basu2021computational}, largely owing to the presence of specific surface receptors that pathogens can exploit for cell invasion, combined with a relatively sparse local mucociliary substrate \cite{lee2019nasopharyngeal}. 
It is also important to note that the nasopharyngeal region contains nasal-associated lymphoid tissue (NALT) \cite{brandtzaeg2011ajrccm,laube2024fdd}, which offers a direct connection to the immune system.
To enhance the therapeutic efficacy against certain pathogens, such as the SARS and influenza viruses, it could be therefore construed essential to improve the targeted delivery of drugs \cite{kashyap2019drug} to the nasopharynx. With that perspective, this study explores the use of intranasal sprays as a method for targeted drug administration to the nasopharynx and models the transport of sprayed drug particulates during relaxed inhalation (at 15 l/min), through experimentally validated computational simulations of the relevant respiratory flow physics inside anatomical domains built from medical imaging. We derive the nasopharyngeal deposition efficiency ($\xi$, in \%) across a broad range of formulation and device parameters, namely the material density of the sprayed formulation ($\rho$, in g/ml), the aerodynamic diameters of the sprayed particles ($d$, in $\mu$m), and the plume angle of the conical spray discharge ($\theta$, in degrees) subtended at the nozzle location by the spray jet. The main takeaways from our study will address the following question: \textit{which specific combination(s) of $d$, $\theta$, and $\rho$ will maximize intra-nasally sprayed drug deposition at the nasopharynx?} To that end, we have derived a three-dimensional parametric diagram (conceptualized in Fig.~\ref{fig1}) whose topological shape can directly identify the conditions for maximal spray performance for targeted delivery at a specific tissue site, such as the nasopharynx.

Maximizing local deposition at infection-prone regions is understandably crucial for improving pharmaceutical effectiveness \cite{foo2007influence,perkins2018ideal,basu2020numerical,tong2016effects}. Traditional methods for optimizing nasal spray formulations and delivery devices often involve trial and error, which can be both time-consuming and costly.
Using full-scale three dimensional computational fluid dynamics (CFD) modeling \cite{feng2021jas,hayati2023ecmf,hosseini2020cbm,dey2025jas,islam2025drug,niegodajew2025flow,basu2021computational,kleinstreuer2008new}, it is however possible to reliably simulate how drug particles behave as they move through the tortuous nasal passages \cite{basu2018computational,kleven2012computational,farnoud2020large}. These models can predict droplet deposition patterns based on factors such as droplet size, spray plume angle, formulation properties, and airflow conditions---thereby offering valuable insights into how to finetune the design of intranasal sprays for better efficacy \cite{inthavong2008cbm}. Herein, we use the same approach to guide the optimization of current formulations along with laying the groundwork for developing CFD-informed augmented intranasal delivery systems. The intra-airway dynamics of the sprayed droplets was modeled by assuming them as inert discrete phase particles bearing appropriate physical properties (in terms of spherical shapes/sizes, material density). Notably, the terminologies `droplets' and `particles' are used equivalently in this paper. 

Systematically pinning down the droplet transport features and the resulting deposition patterns within realistic nasal cavities is crucial toward designing new-generation sprays that can effectively target the disease-prone tissue regions along the airway. The findings reported here build on our previous publication in this journal \cite{akash2023model}. While the prior study had focused on refining the spray axis orientation and nozzle position within the anterior respiratory airspace, the current work invokes the same spray placement protocol but provides significant updates on improving device and formulation design to maximize targeted drug delivery at the nasopharynx. Preliminary results from this study have been presented at the 2023 Annual Meeting of the American Physical Society's Division of Fluid Dynamics \cite{malakar2023aps}.


\begin{figure}[t!]
\begin{center}
\includegraphics[width=\textwidth]{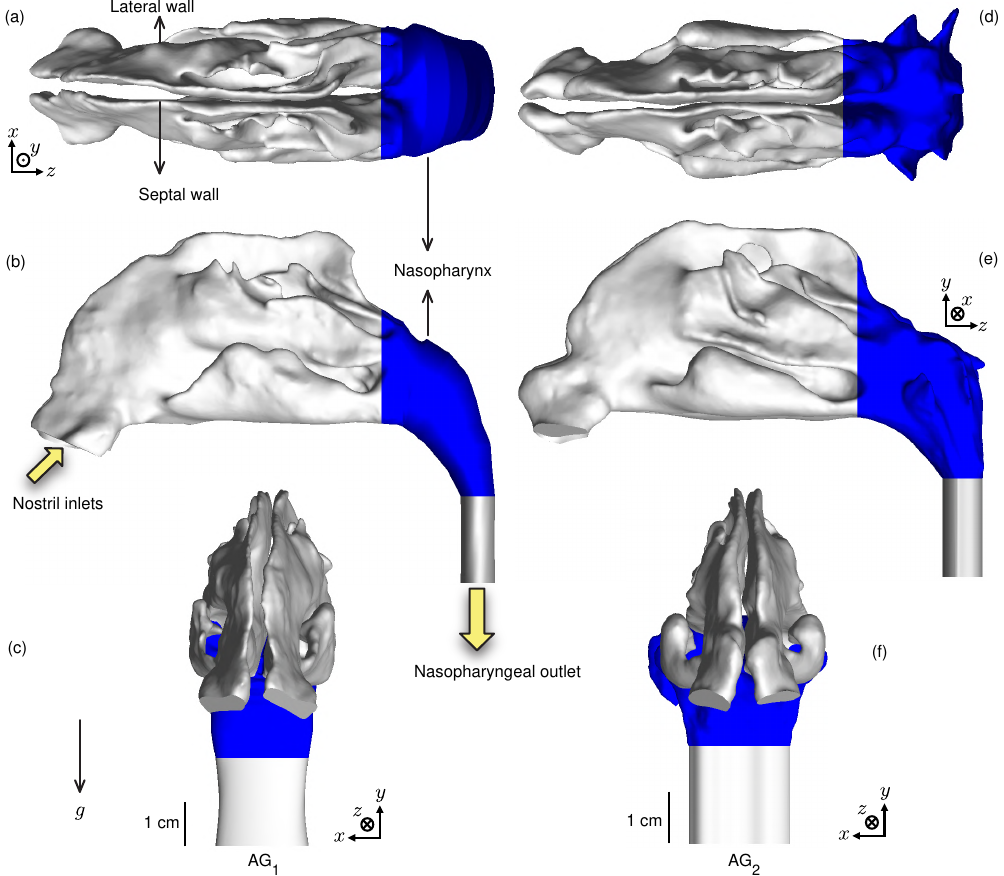}
\caption{{\bf Test upper airway geometries:} Panels (a)-(c) respectively show the axial, sagittal, and coronal views of the computed tomography (CT)-based anatomical reconstruction of AG$_1$ (anatomical geometry 1). Similarly, panels (d)-(f) respectively show the axial, sagittal and coronal views of the CT-based reconstruction of AG$_2$ (anatomical geometry 2). The nasopharynx is colored blue in the visuals \cite{borojeni2017ijnmbe}. The solid yellow arrows indicate the airflow inlet and outlet regions for the inhalation simulations, with their directions aligning with the direction of mean flux. $g$ implies the gravity direction in the simulations. Additionally, the visuals in (c) and (f) show the geometry-specific length scales. (a)-(c) have the same scale and (d)-(f) have the same scale.}
\label{fig2}
\end{center}
\vspace{-0.5cm}
\end{figure}


\section{Materials and Methods}\label{methods}

\subsection{Anatomical domain reconstruction}\label{subjects}

The anatomical upper airway geometries (see Fig. \ref{fig2}), used in this study, were rebuilt from existing, de-identified, medical-grade computed tomography (CT) imaging data collected from two adult test subjects with disease-free airways. Therein, the coronal depth increments in between the CT slices were $\approx$ 0.4 mm. We have named the test domains as anatomic geometry 1 (or, AG$_1$) and anatomic geometry 2 (or, AG$_2$).
The computational retrospective use of the existing, anonymized scans was approved with exempt status by the Institutional Review Board at South Dakota State University. For anatomical precision, the nasal airspace segmentation was carefully hand-edited after applying the expected radio-density delineation of -1024 to -300 Hounsfield units \cite{basu2018computational}. For this step, the high-resolution DICOM (Digital Imaging and Communications in Medicine) scans for each subject were imported into the image-processing program Mimics Research 18.0 (Materialise, Plymouth, MI).

Subsequently, we imported the rebuilt geometries to ICEM CFD 2024 R1 (ANSYS Inc., Canonsburg, Pennsylvania) as stereolithography (STL) files. To spatially mesh the reconstructed cavities according to established mesh refinement-based protocols \cite{frank2016influence,basu2017influence}, each computational grid included 3 prism layers ($\approx$ 0.1 mm thick) along the airway walls (except the nostril inlet planes and the outlet plane) with a height ratio of 1. For the core cavity space, approximately 4.2 million (in AG$_1$) and 4.4 million (in AG$_2$) unstructured, tetrahedral elements were generated by implementing the tetra/mixed type mesh with robust (Octree) method. Combined with the prismatic element counts, the total element numbers were 5.3 million in AG$_1$ and 5.4 million in AG$_2$.

\noindent\textit{Spray axis determination and nozzle placement:} The spray placement in the digitized airspace domains followed the `line-of-sight' (LoS) protocol established by us previously \cite{basu2020numerical,akash2023model,treat2020ro} for improved drug delivery, whereby the spray axis should (virtually) cut through the target tissue site. Accordingly, after ascertaining the centroid of the nostril plane (through which spray would be administered) in each reconstructed geometry, we identified an arbitrary point generally positioned near the upper edge of the nasopharynx. The direction vector between the nostril centroid and the located point provided a repeatable spray direction, with the spray nozzle placed 5-mm into the airspace from the nostril centroid. Note that the nasopharynx comprises the upper segment of the pharynx at the back of the nose, after the two sides of the anterior nasal airspace merge; see Fig.~\ref{fig2}.

\begin{figure}[t!]
\centering
\includegraphics[width=\textwidth]{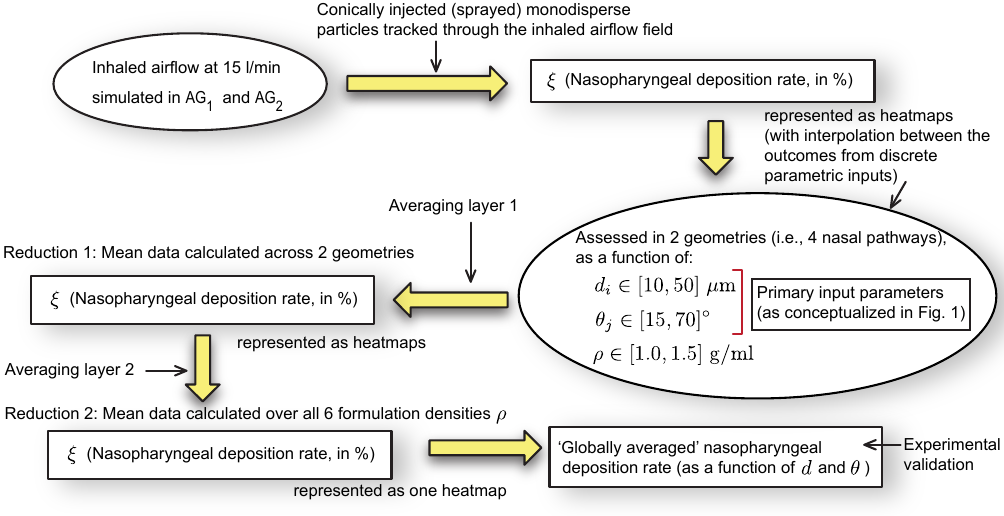}
\caption{{\bf Schematic workflow:} This computational study assesses the nasopharyngeal deposition rates ($\xi$) for discrete combinations of formulation densities ($\rho$), plume angles ($\theta$), and sprayed particle diameters ($d$). The deposition heatmaps (included in the results) give interpolated visuals from the discrete assessments. As per the schematics in Fig.~\ref{fig1}, the input parameter $i$ and input parameter $j$ are $d$ and $\theta$ (in no specific order). We generate $\xi$ for a wide range of $d$ and $\theta$, for six specific $\rho$. Eventually, the `globally averaged' $\xi$ is obtained, by averaging the nasopharyngeal deposition assessments for all $\rho$ and test geometries (with two layers of averaging, as illustrated above).}
\label{fig3}
\end{figure}

\subsection{Numerical simulations}\label{s:methods}
\subsubsection{Inhaled airflow and sprayed particle tracking simulations}

This study investigated the intra-airway deposition behavior of 3,000 monodisperse particles---each set bearing aerodynamic diameters  $d \in [10,50]$ $\mu$m (with increments of 1 $\mu$m). These particles were sprayed into the anatomical domains carrying an underlying airflow field that mimicked resting inhalation rate of 15 l/min \cite{garcia2009dosimetry,borojeni2020ijcars}. The exercise was performed for six formulation densities, $\rho \in [1.0, 1.5]$ g/ml (with increments of 0.1 g/ml) and twelve plume angles, $\theta \in [15\degree,70\degree]$ (with increments of 5$\degree$); see Figs.~\ref{fig2} and \ref{fig3} for the conceptualized study design. A cone injection technique was employed to ensure realistic drug delivery and reliable particle deposition measurements.

The inhalation airflow was modeled using the Large Eddy Simulation (LES) scheme that resolved turbulent flow structures, dividing the turbulence \cite{longest2007jb,doorly2008rpn} into large- and small-scale motions. Subgrid-scale kinetic energy transport model was invoked to track small fluctuations \cite{baghernezhad2010different,farnoud2020large}. The computational simulations were performed on ANSYS Fluent 2024 R1, with the implementation of a segregated solver. Therein we used the SIMPLEC pressure-velocity coupling and second-order upwind spatial discretization. The solution convergence was monitored by minimizing the mass continuity residuals to $10^{-2}$ and velocity component residuals to $10^{-6}$. Additionally, the stabilization of mass flow rate and static pressure at airflow outlets, namely at the nasopharyngeal outlet (see Fig.~\ref{fig2}), was kept track of. 
For these pressure gradient-driven airflow solutions, the LES work necessitated a computing time of 1.5 to 2 days, to replicate inhalation flow over a span of 0.35 seconds (adequate for sprayed particle transport to cover the nostril-to-nasopharynx pathway), using a time-step of 0.0002 seconds \cite{ghahramani2017cf}. To accurately mimic warmed-up air inside the respiratory route, its dynamic viscosity was set at $1.825\times10^{-5}$ kg/m.s and the density was assumed as 1.204 kg/m$^3$.

\begin{figure}[t!]
\centering
\includegraphics[width=\textwidth]{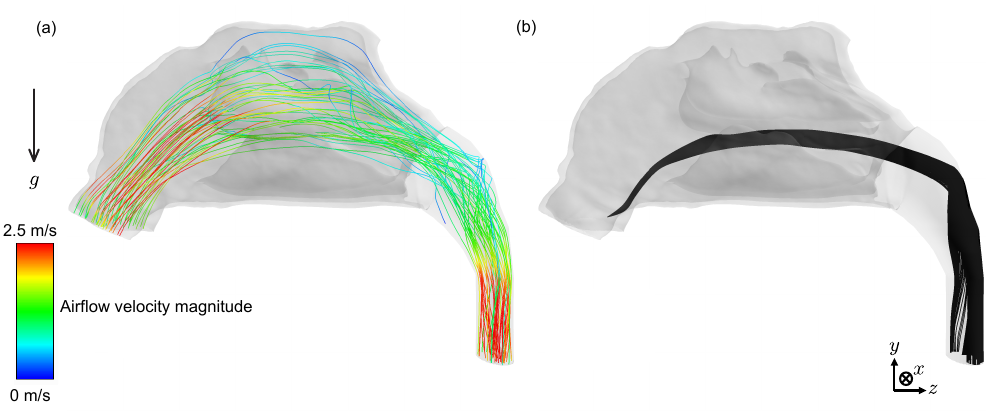}
\caption{{\bf Representative flow field and sprayed particle trajectories:} (a) Sample airflow velocity streamlines within AG$_1$ mimicking 15 l/min inhalation. These representative streamlines initiate from 25 random points on each nostril (i.e., a total of 50 streamlines are shown above). (b) Sample spatial trajectories of intra-nasally sprayed particles, with $\rho =$ 1.0 g/ml, $\theta =$ 15\degree, and $d =$ 18~$\mu$m. }
\label{fig4}
\end{figure}

The tracking of intranasal spray dynamics against the surrounding inhaled airflow was accomplished using Lagrangian-based inert discrete phase simulations (e.g., see Fig. \ref{fig4}) with Runge-Kutta solver. The motion of the sprayed particles was assumed to be one-way coupled with the surrounding flow \cite{inthavong2008cbm,feng2017bio,zhao2021pof}, meaning that the particles' trajectories were influenced by the flow features, but they did not, in turn, affect the airflow field around them. The simulations integrated the particle transport equation that accounted for various forces acting on small particulates, such as the ambient inhaled airflow drag, gravity, and other appreciable body forces (namely the Saffman lift force relevant for small particles). While deriving the particle deposition data, we implemented a no-slip trap boundary condition on the walls of the cavity, enabling the assessment of localized droplet clustering over intranasal tissues. For each formulation density, the sprayed droplets (also often referred to as `particles' in this study) were introduced into the airspace as a solid-cone injection starting from the nozzle point. The initial velocity of the droplets was realistically set at 10 m/s \cite{liu2011assessment} and a total non-zero mass flow rate of $1\times10^{-20}$ kg/s was given as the initial condition of the streams in the spray cone.

The following expands on the boundary conditions during particle tracking: (i) The airway-tissue interface, which represented the walls enclosing the digitized nasal airspace, had a zero tangential velocity condition (commonly known as the `no-slip' condition); additionally, the walls were enforced with the `trap' discrete phase boundary condition, enabling the particle tracks to cease once they enter the elements adjoining the walls. (ii) For the nostril planes, a `reflect' discrete phase boundary condition was used to simulate the effect of inhalation on the particle trajectories if they were on the verge of falling out of the anterior nasal domain. (iii) The airflow outlet plane, designated as the pressure-outlet zone, had an `escape' discrete phase boundary condition, allowing the outgoing particle trajectories to exit the upper respiratory airspace. Considering the area-weighted average of the inlet and outlet pressure variables in the simulations, the mean total pressure gradient driving the 15 l/min airflow in the two test geometries was 5.63 Pa (with a strikingly comparable 5.66 Pa in AG$_1$ and 5.59 Pa in AG$_2$).

For details on the mathematical framework of the numerical scheme employed in this study, please refer to \cite{basu2024arXivURTLRT}. The computational approach has also been thoroughly validated in one of our earlier publications \cite{basu2020numerical}. This validation involved comparing the regional deposition patterns along the inner walls of \textit{in silico} nasal anatomical models with gamma scintigraphy measurements of regional deposition obtained from \textit{in vitro} spray tests conducted on 3D-printed solid transparent replicas with similar reconstructions.

\begin{figure}[t!]
\centering
\includegraphics[width=1\linewidth]{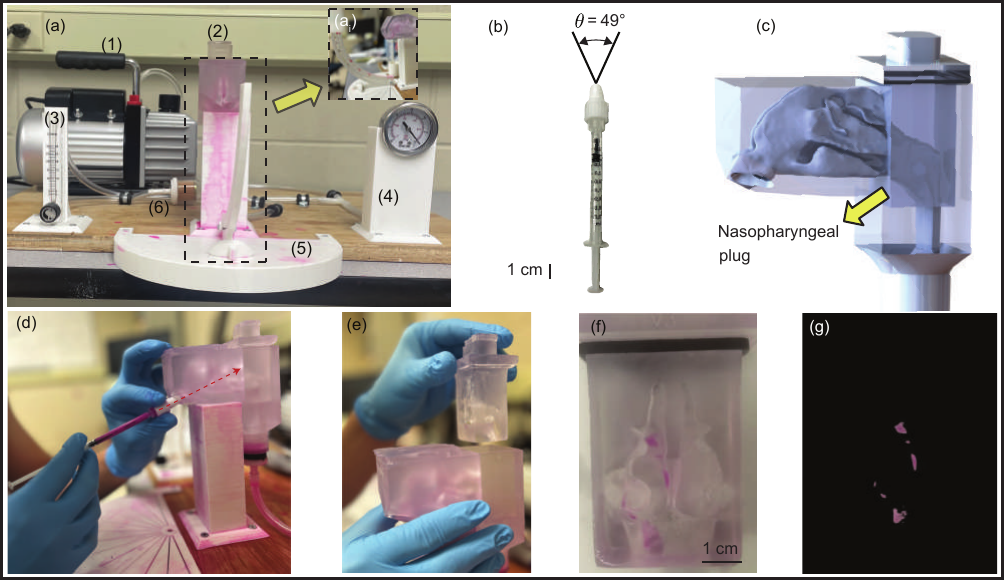}
\caption{{\bf Setup for experimental validation:} Panel (a) shows the front view of the experimental setup, comprising the following numbered components: (1) a 2.5 CFM vacuum pump, (2) a 3D-printed nasal airway model, (3) a flow rate meter, (4) a pressure gauge, (5) a spray cone angle indicator, and (6) an air filter. Inset $a_i$ shows the side view of the setup. Panel (b) demonstrates the LuerVax\textsuperscript{TM} spray device (an Aptar Pharma product) used in the experiments. The spray plume angle is $\theta = 49\degree$. Panel (c) presents the 3D CAD visual of the nasal airway cast, with the nasopharyngeal region constructed as a removable plug system for ease of measuring the local deposits. See Fig.~\ref{fig2}f, for the length scale of the life-size anatomical reconstruction.
Panel (d) illustrates the procedure of the experimental spray trials, with the dotted red arrow representing the spray axis cutting through the nasopharynx. Panel (e) shows the removable nasopharyngeal plug. Panel (f) provides a representative image taken after one of the spray trials, while panel (g) shows the corresponding image-processed view of (f).} 
\label{fig5}
\end{figure}

\subsection{Experimental setup}\label{experimentalmethods}

With AG$_2$ demonstrably exhibiting higher nasopharyngeal deposition (per the simulations; see \S\ref{s:results}, Figs.~\ref{fig6} and \ref{fig7}), a 3D-printed cast of AG$_2$ was built for physical experimental tests,  with sprays administered through its both nasal openings. The experimental setup (see Fig.~\ref{fig5}a) included a 2.5 cubic feet per minute (CFM) vacuum pump (Pittsburgh Automotive\textregistered, 120VAC / 60Hz / 3.2A), a flow rate meter connected to an air filter, a pressure gauge, a stable platform for securing the geometry model and a spray cone angle indicator. The experiments were conducted using LuerVax\textsuperscript{TM} spray device (see Fig.~\ref{fig5}b and d) filled with a fluorescent dye solution diluted in distilled water, with the solution having an approximate density of 1.054 g/ml and the spray administered with a measured plume angle of $\theta =$ 49$\degree$ (see Fig.~\ref{fig5}b; also see Table~\ref{table1}). 

The 3D geometry of AG$_2$ was printed in a stereolithography (SLA) 3D printer, Anycubic Photon M3 Max, by employing high-clear resin in an attempt to yield optical transparency. The optical transparency of the resulting resin was critical, as it enabled visualization of internal flow structures as well as post-deposition patterns after fluorescent excitation. SLA printing  was chosen over other forms of additive manufacturing \cite{he2019ijamt}---given its ability to yield higher resolution, smooth finish over the whole surface, as well as higher fidelity in replicating complex internal anatomical structures.

In addition, the nasopharyngeal portion (see Fig.~\ref{fig2}) was fabricated in the form of a removable plug (shown in Fig.~\ref{fig5}c-f), and instead of re-using the same plug, 20 different plugs were used during 10 spray trials run through each nostril. The spray administration protocol (comprising 10 trials in each nasal opening; the nozzle being $\approx$ 5-mm into the airspace along the LoS direction) was carefully tuned for a robust quantification of the deposited mass at the nasopharynx. The use of clear resin also minimized scattering of radiation upon imaging, therefore further permitting more reliable as well as reproducible quantification of deposited droplets. Following each individual test, high-resolution images of the anterior, posterior, and lateral portions of the plug were captured. These images were processed in MATLAB using color masks (e.g., see Fig.~\ref{fig5}f-g) to generate a percentage of deposition relative to the image. Each trial had the sum of its depositions calculated from MATLAB to quantify dye deposition within the nasopharyngeal plug.


\section{Results}\label{s:results}

\subsection{Numerical simulation results}
\subsubsection{Variation trend in nasopharyngeal deposition}

\begin{figure}[H]
\centering
\includegraphics[width=\textwidth]{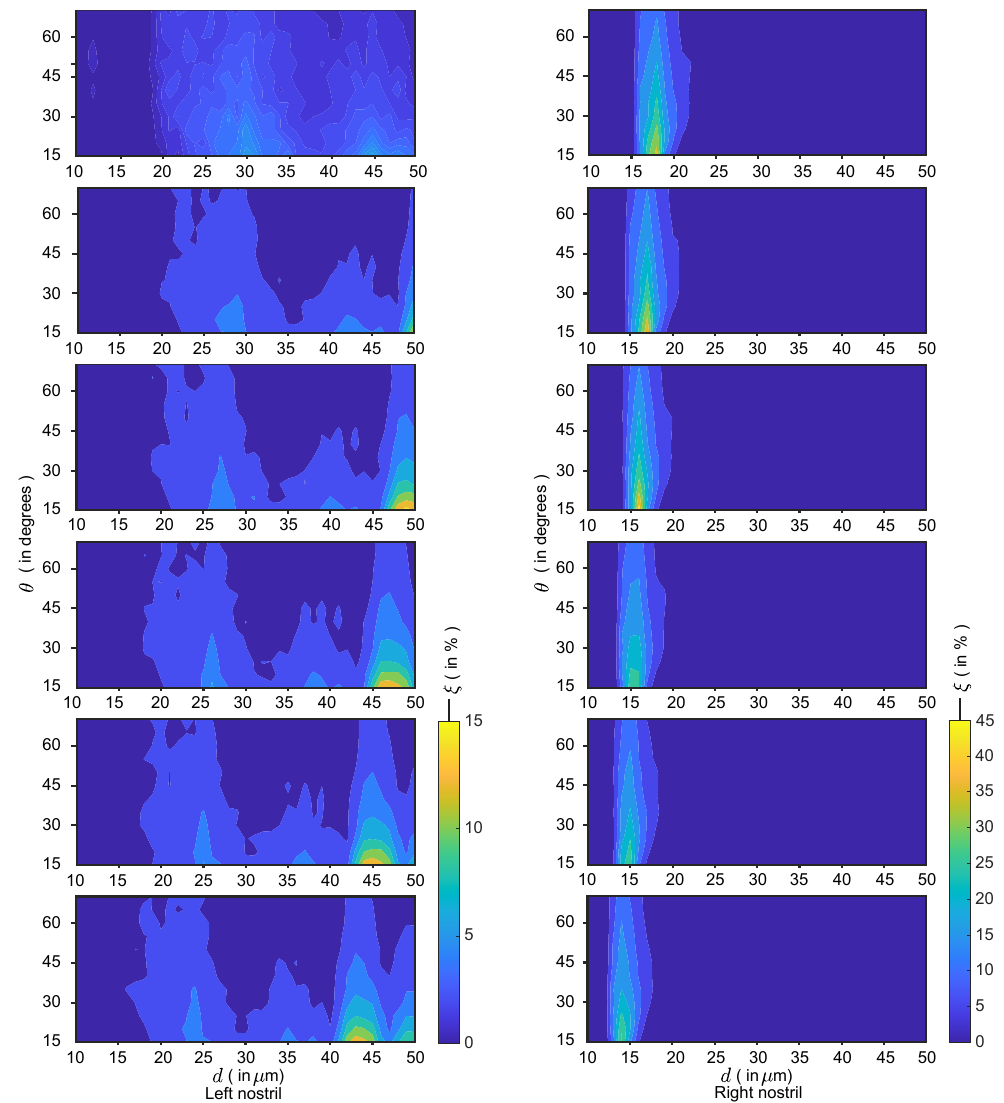}
\caption{{\bf Simulated nasopharyngeal deposition trend for AG$_1$:} Contour plots for nasopharyngeal deposition rate (as in, what fraction of the monodisperse particles end up depositing at the nasopharynx; represented as $\xi$) as a function of the spray plume angles ($\theta$, recorded along the horizontal axis) and particle diameters ($d$, recorded along the vertical axis). Owing to the effect of inertial impaction on downwind penetration, the optimal parametric region (for maximal $\xi$) gradually shifts toward the left side of the contour map with increasing formulation density $\rho$. Top-to-bottom: $\rho \in [1.0,1.5]$ g/ml, with an increment of 0.1 g/ml in successive rows. Left column: left nostril data; right column: right nostril data.}
\label{fig6}
\end{figure}

\begin{figure}[H]
\centering
\includegraphics[width=\textwidth]{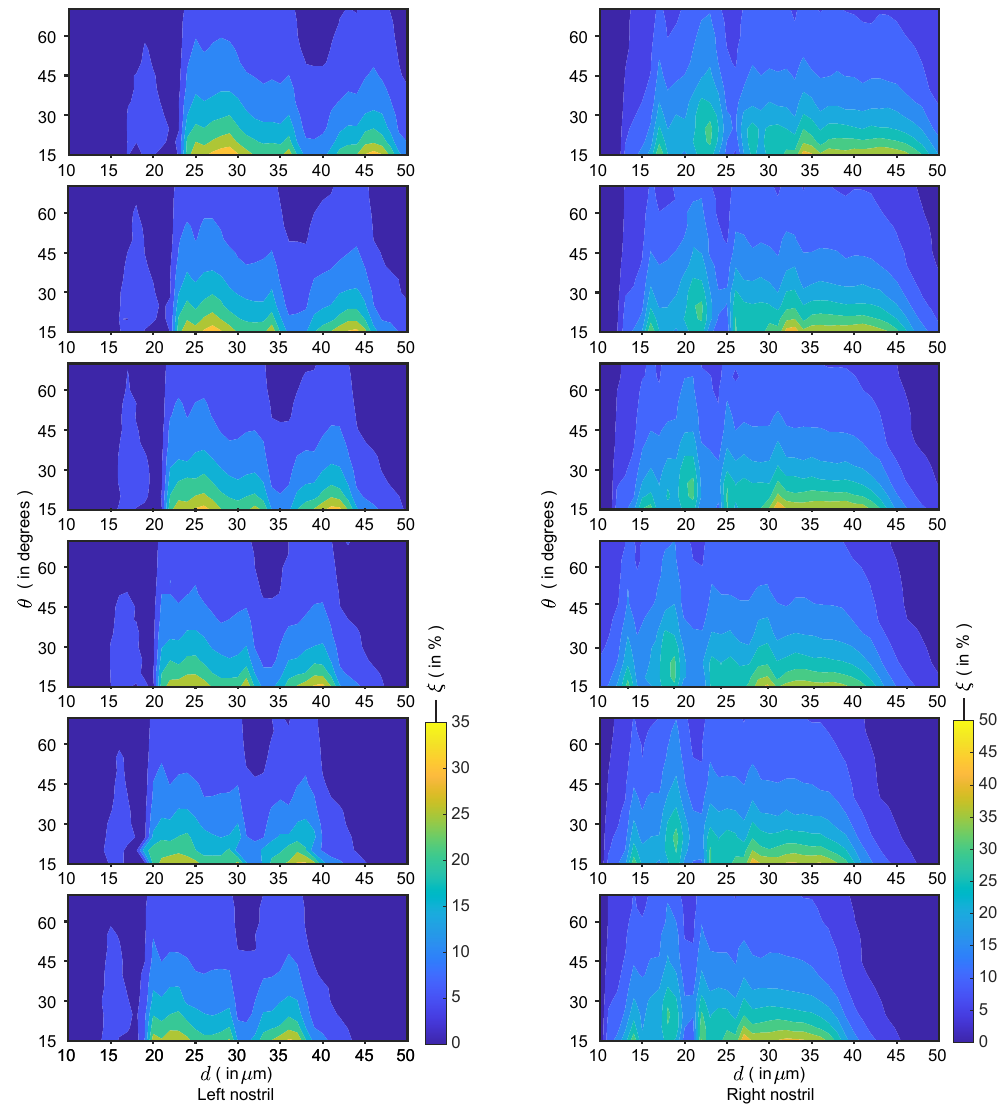}
\caption{{\bf Simulated nasopharyngeal deposition trend for AG$_2$:}  Contour plots for nasopharyngeal deposition rate (as in, what fraction of the monodisperse particles end up depositing at the nasopharynx; represented as $\xi$) as a function of the spray plume angles ($\theta$, recorded along the horizontal axis) and particle diameters ($d$, recorded along the vertical axis). Owing to the effect of inertial impaction on downwind penetration, the optimal parametric region (for maximal $\xi$) gradually shifts toward the left side of the contour map with increasing formulation density $\rho$. Top-to-bottom: $\rho \in [1.0,1.5]$ g/ml, with an increment of 0.1 g/ml in successive rows. Left column: left nostril data; right column: right nostril data.}
\label{fig7}
\end{figure}

\begin{figure}[H]
\centering
\includegraphics[width=\textwidth]{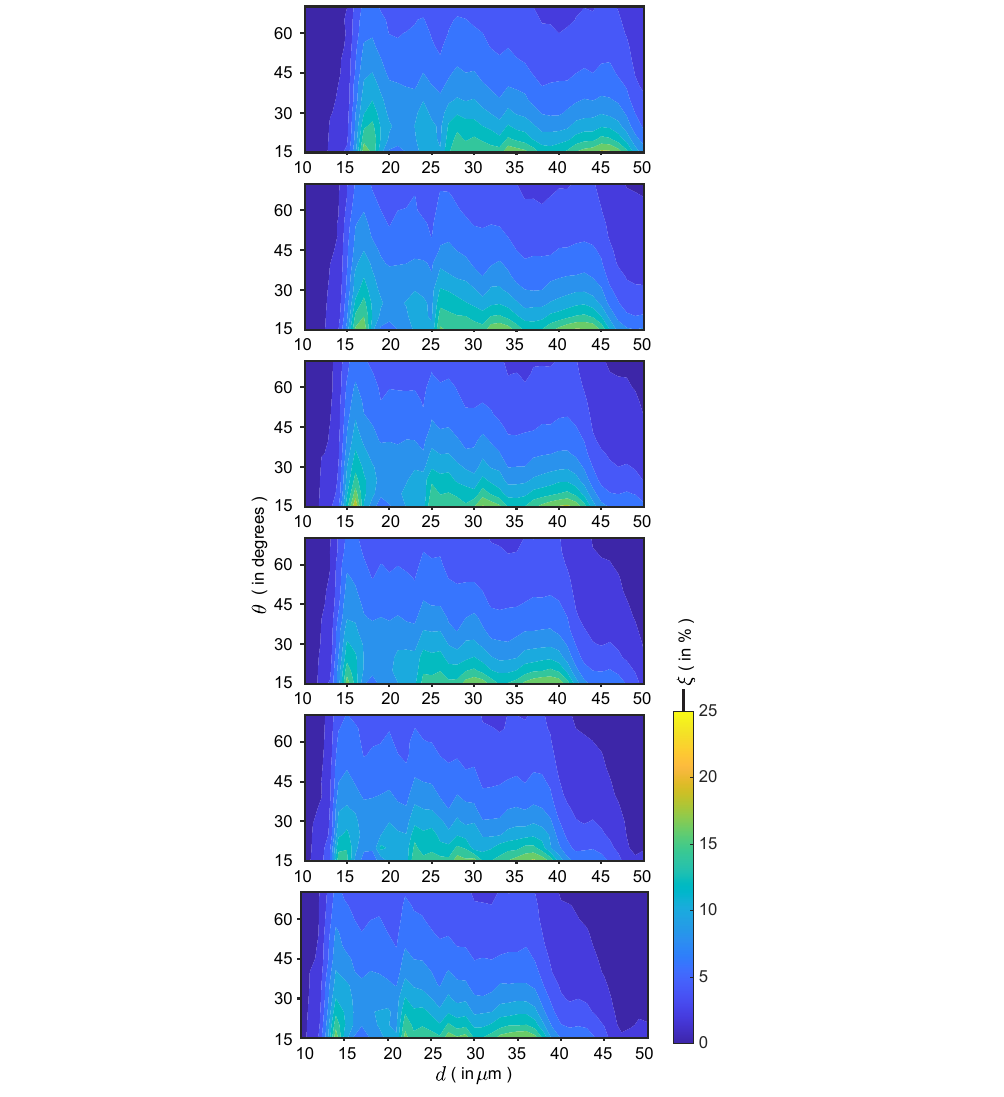}
\caption{{\bf Mean simulated deposition rates for each test formulation density:} Contour maps for mean nasopharyngeal deposition rate ($\xi$) for each test formulation density, $\rho$, which varies from 1.0 g/ml to 1.5 g/ml (with increments of 0.1 g/ml), depicted progressively from top panel to the bottom panel. The averaging considers both nasal sides of AG$_1$ and AG$_2$. 
}
\label{fig8}
\end{figure}

In our study, six different formulation densities were used, resulting in a total of 24 individual contour plots illustrating simulated nasopharyngeal deposition rates ($\xi$, in \%) across the tested airways (see Figs.~\ref{fig6} and \ref{fig7}). Comparing the panels top-to-bottom, the variations (note the left-ward shift) are primarily owing to the particles becoming inertially stronger with increasing formulation density. Also, the topological differences on the contour maps along each row in both Figs.\ref{fig6} and \ref{fig7} indicate the subtle dependence of the local deposition trend on anatomy-specific geometric features. Next, moving beyond the geometric subjectivity, Fig.~\ref{fig8} maps $\xi$ when averaged across all four airspaces, for each test formulation density ($\rho\in[1.0,1.5]$ g/ml, with increments of 0.1 g/ml on panels therein from top-to-bottom).

To expand further, focusing on each column of panels in the contour maps (i.e., for data from the same airway as $\rho$ is progressively raised), we observe that the plots gradually shifted to the left side because of the inertial effect of the formulation densities (higher the density of the solution, higher would be the inertia).  Also, as shown in Figs.~\ref{fig6} and \ref{fig7}, as higher densities increase the inertial effects, particles of similar sizes result in less deposition at the nasopharynx. This behavior can be explained through the Stokes number $Stk =\rho d^2 U/18 \mu D$ (not considering slip correction), where $\rho$ = particle material density, $d$ = particle diameter, $U$ = characteristic fluid speed, $\mu$ = viscosity coefficient of the underlying fluid (air), and $D$ = characteristic cavity diameter . Physically, $Stk$ represents the ratio between local transient inertia and viscosity. Increase in $\rho$ enhances the inertial dominance in particle motion. Everything else (including the diameters) staying same---the higher inertial particles are more likely to settle in the front sections of the nose owing to inertial impaction (coupled with gravitational sedimentation), rather than penetrating further into the complex nasal passage to reach the nasopharynx. Conversely, particles with $Stk \leq$  1 are carried more efficiently downwind by the fluid streamlines they are embedded in \cite{aggarwal1994effect,dey2025improved,schroeter2011jas,finlay2001book}.

\newpage

\begin{figure}[H]
\centering
\includegraphics[width=\textwidth]{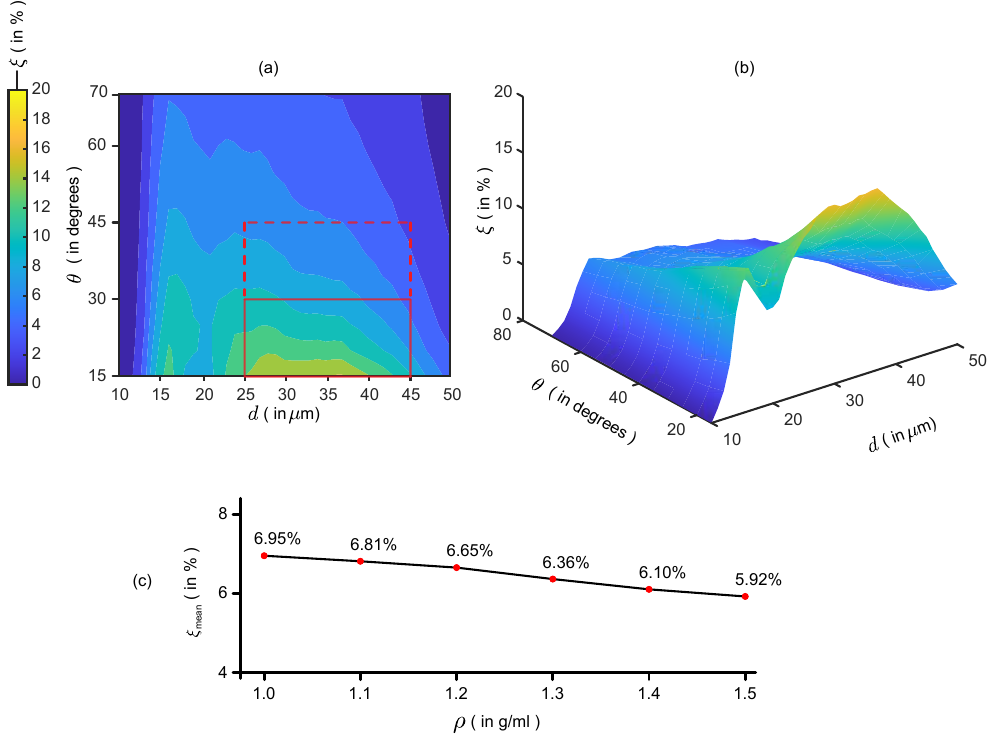}
\caption{ {\bf Optimal parametric choices in AG$_1$ and AG$_2$:} Panels (a) and (b) depict the `globally averaged' contour map for nasopharyngeal deposition rate ($\xi$, in \%), as the input parameters $d$ and $\theta$ are varied. Panel (b) is the isometric view of the same data shown in (a). The simulated data here is obtained by averaging all the panels from Figs.~\ref{fig6} and \ref{fig7}, considering all the test airways and formulation densities. 
The rectangle bounded by solid red lines in (a) demarcate the parametric choices $d \in [25,45]$ $\mu$m and $\theta \leq 30\degree$, when $\xi$ visibly peaks. The mean $\xi$ is 11.37\% within this region. If we extend the vertical bound to $\theta \leq 45\degree$ (marked by the dashed red lines), the mean $\xi$ reduces slightly to 9.38\%.  
Compare these to the mean deposition rate for each of the formulation densities throughout the $(d,\theta)$ domain, as shown in panel (c). The average from the 6 recorded means is 6.47\%. \textbf{Note:} The reader may find it intriguing to compare the visual in panel (b) to the hypothesized conceptual diagram pitched in Fig.~\ref{fig1}.}
\label{fig9}
\end{figure}

\begin{figure}[H]
\centering
\includegraphics[width=\textwidth]{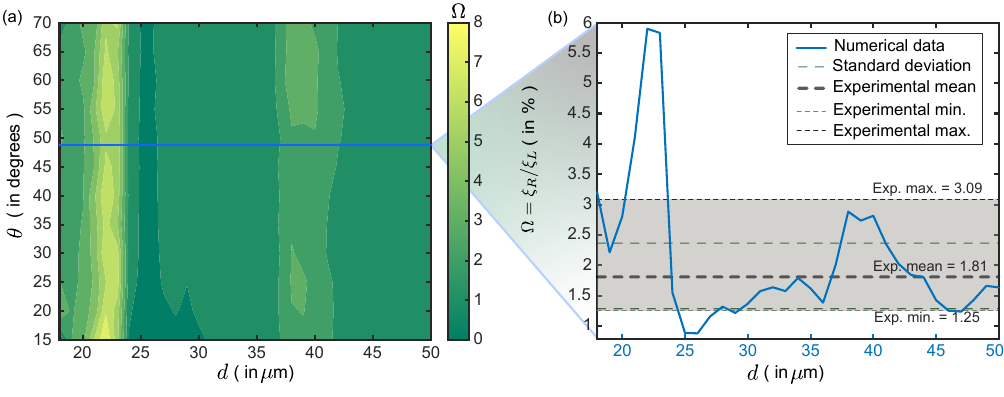}
\caption{{\bf Comparison between numerical and experimental test cases:} Panel (a) presents the contour plot obtained from numerical simulations showing the ratio of right ($\xi_R$) -- to -- left ($\xi_L$) nostril deposition efficiencies (symbolically represented as $\Omega$) at nasopharynx for particles with a material density of 1.0 g/ml, corresponding to the experimental condition (with the sprayed formulation particulates comprising water embedded with fluorescent markers and bearing a material density of $\approx$ 1.05 g/ml). The blue horizontal line (in panel (a)) marks the simulation data for the plume angle of 49$\degree$, which is the measured $\theta$ for the LuerVax\textsuperscript{TM} nozzle (an Aptar Pharma product) used in the experiments. Using the blue line, panel (b) illustrates the simulated variation of the right-to-left nostril deposition ratio (at nasopharynx) for the test particle sizes for this specific plume angle. The shaded light-gray region denotes the experimentally measured right-to-left deposition ratio range. The thick gray dashed line indicates the experimental mean which is 1.81, the upper and lower light dotted lines respectively represent the maximum (3.09) and minimum values (1.25) from the experimental trials, and the light dashed green lines denote the standard deviation on the experimental $\Omega$.
The reader should note that the gray horizontal band in (b) represents data from physical experiments conducted with a realistic droplet size distribution in each actuation. It does not correspond directly to the $d$ values (shown in blue) along the horizontal axis. In contrast, the simulation-derived $\Omega$ (blue plot) are functions of $d$, since the simulations assume monodisperse droplets. It is reassuring that the simulation results and experimental data agree when $d \gtrsim 25~\mu$m, as the experimental spray shots are intentionally composed of particles dominated by that size range (e.g., refer to Table~\ref{table1}).
} 
\label{fig10}
\end{figure}

\subsubsection{\bf Generic parametric bounds for enhanced $\xi$}
Figure~\ref{fig9}a-b show the nasopharyngeal deposition rate ($\xi$, in \%) when `globally' averaged across all the test airway domains and formulation densities; the flowchart for the averaging algorithm is shown in Fig.~\ref{fig3}. Within the red rectangle in Fig.~\ref{fig9}a (where $\xi$ visually peaks), the maximum and minimum $\xi$ are respectively 16.24\% and 5.55\%. The mean deposition rate within the same parametric bounds with $d\in[25,45]$ $\mu$m and $\theta\leq30\degree$ (within the red solid line-bounded rectangle) is 11.37\% (averaged from the $\xi$ outcomes for the discrete $(d,\theta)$ inputs within the said domain). The latter is 76\% higher than the global mean $\xi$ across the entire parametric domain (see Fig.~\ref{fig9}c). 
It is worth noting that these particles within the red rectangle are noticeably larger than the typical aerosol dimension of 5~$\mu$m. Particles smaller than 10~$\mu$m would predominantly bypass the nose, penetrating downwind to deposit in the laryngotracheal cavity and the bronchial recesses \cite{crowder2002fundamental,chakravarty2022pulmonary,darquenne2022aerosol,basu2024arXivURTLRT}. From a regulatory and manufacturing perspective in context to designing optimized spray formulations for upper airway target sites (such as the nasopharynx), the findings will thus mitigate a potential caveat, as the US Food and Drug Administration (FDA) diligently monitors the proportion of droplets measuring less than 10 $\mu$m to ensure safety.

\subsubsection{\bf Impact of spray plume angle $\theta$}
The geometric features of the spray delivery system are crucial for directing particles to the desired intra-airway locations. The plume angle $\theta$ is the solid cone angle subtended at the delivery device nozzle that sprays the pharmaceutical agents (e.g., see Fig.~\ref{fig5}b). We have investigated 12 different plume angles for both geometries ranging from 15\degree to 70\degree (with an increment of 5\degree). Figs.~\ref{fig6} -- \ref{fig8} show that there is, in general, an inverse relation between the plume angles ($\theta$) and nasopharyngeal deposition rates ($\xi$). Higher plume angles result in lower deposition $\forall \rho$,  in both subjects (AG$_1$ and AG$_2$). This is expected as a narrower, more pin-pointed spray profile can understandably deliver drugs more efficiently downwind, given the spray axis is directed aptly at the region of interest. Enforcing a threshold of around 5\% deposition, it is observed that $\theta\leq60\degree$ can yield substantial results for particle sizes falling within the 20 – 45 $\mu$m range. A more effective choice is $\theta\leq30\degree$, which (as shown in Fig.~\ref{fig9}a) leads to a mean $\xi$ that exceeds 11\%, for $d\in[25,45]$ $\mu$m.

\subsection{Representative experimental validation}
Physical experiments comprising nasal spray administration (mimicking the computational spray delivery protocol, with LuerVax\textsuperscript{TM}) were performed 10 times per nostril within the 3D printed cast of AG$_2$. Each test consisted of an airflow rate of 15 l/min passing through the anatomical cast (mimicking inhalation; e.g., see Fig.~\ref{fig5}) and ten pumps of diluted dye from the spray bottle inserted into the printed model (see \S\ref{experimentalmethods} for details on the sprayed solution). For each trial, the deposition rates for the right nostril are found higher than those for the left nostril; this trend is in agreement with the numerical results for AG$_2$ (see Figs.~\ref{fig7} and \ref{fig10}a). Each trial then compared the respective nasopharyngeal depositions of the two nostrils to generate a ratio. During the ten trials, the ratio (represented as $\Omega$) of the experimental $\xi$ when sprayed through the right nostril to the $\xi$ when sprayed through the left nostril---averaged 1.81, with the  minimum and maximum values being, respectively, 1.25 and 3.09; see Fig.~\ref{fig10}b. $\Omega$ values from the simulated data are plotted as the blue line in Fig.~\ref{fig10}b. The values correspond to $\theta = 49\degree$ in the simulations, it being the measured plume angle for the spray device used in the experiments. The simulations and experiments agree when $d$ is approximately within $[28,50)$ $\mu$m. 

Herein (i.e., Fig.~\ref{fig10}b), note that the gray horizontal band (and the gray/green dashed bars) present data from physical experiments conducted with a realistic droplet size distribution in each actuation and do \textit{not} have any correlation to the $d$ values (in blue) along the horizontal axis in Fig.~\ref{fig10}b. In contrast, the $\Omega$ values along the simulation-derived blue plot are functions of $d$ (since the simulations were monodisperse). Therefore, it is reassuring that the simulations and experiments align when $d \gtrsim 25$~$\mu$m, as the experimental spray shots (by design) are predominantly composed of particles within that size spectrum (see Table~\ref{table1}).


\begin{table}[t!]
\centering
\renewcommand{\arraystretch}{1.5} 
\setlength{\tabcolsep}{12pt} 
\begin{tabular}{lccccc}
\hline
\textbf{Product} & \textbf{Dv$_{10}$} ($\mu$m) & \textbf{Dv$_{50}$} ($\mu$m) & \textbf{Dv$_{75}$} ($\mu$m) & \textbf{Dv$_{90}$} ($\mu$m) & $\theta$ (degrees) \\
\hline
BiVax 200\textsuperscript{TM} & 19 & 36 & 49 & 64 & 69 $\pm$ 2 \\
LuerVax\textsuperscript{TM} & 20 & 43 & 64 & 89 & 49 $\pm$ 1\\
\hline
\end{tabular}
\caption{Mean droplet size distribution (DSD) and plume angles ($\theta$) in Aptar Pharma spray products for intranasal vaccination (targeting the nasopharynx). Errors on $\theta$ are a measure of the corresponding standard deviations in the measurements (with 5 trials for each product). The DSD was determined via laser diffraction using a Malvern
Spraytec\textsuperscript{\textregistered} (Malvern Instruments, Worcestershire, United Kingdom). $\theta$ was evaluated using a SprayVIEW\textsuperscript{\textregistered} (Proveris Scientific; Hudson, MA) measuring system, which is a non-impaction laser sheet-based instrument. For the listed measurements, the atomizer was positioned 5 cm from the laser for plume angle measurements and 4 cm from laser for the DSD measurements \cite{baxter2022aaps}. Detailed parameters associated with the measurement techniques are available in \cite{laube2024fdd}.}
\label{table1}
\end{table}

\section{Discussion}

\subsection{Perspectives on the modeling approach}
Despite the potentially valuable insights gained regarding the optimal design for targeted nasal sprays, this study can however be critiqued for several limitations that relate back to the true biological realism of the investigative framework. An important limitation is the assumption of structural rigidity in the nasal anatomical reconstructions. Although the geometries were built with high fidelity from medical-grade imaging, they do not factor in the temporally dynamic, elastic properties of nasal tissues, which can significantly influence local airflow patterns and particle deposition under physiological conditions. Integrating tissue compliance and mucosal movements could give a more precise estimate of drug delivery at targeted disease-prone sites along the intra-airway tissue lining.

Another key nuance is the absence of particle size distribution consideration in the (monodisperse) simulations. Actual spray products offer a heterogeneous size distributions of aerosols and microdroplets. However, this study was designed to systematically identify which specific particle sizes are most suited at \textit{directly} reaching the nasopharynx through the spraying action; (it is expected that) the information could then guide the design of real sprays with their particle sizes geared toward the precise findings of this study.

Next, we have overlooked (for now) the potential chemical and/or biological interactions within the nasal mucosa, such as mucociliary clearance or enzymatic activity, which can impact deposition (and therapeutic) efficacy over time. Another somewhat-related and crucial consideration involves the toxicological safety associated with increased targeted deposition. While larger particles like the ones between 25 – 45 $\mu$m are demonstrably likely to directly deposit at the nasopharynx (through spraying) and thus effectively deliver the therapeutic agents, the size of the particles and the material density of the formulation could have implications for safety profiles. Particles within specific size ranges may present risks of localized irritation or trigger immune responses, and the formulation properties may necessitate comprehensive toxicological assessment to avert adverse effects, including inflammation or unintended tissue damage. For the latter, it should however be noted that all other parameters remaining same, the 1 g/ml water-like formulation constitution did give the highest mean nasopharyngeal deposition.

Finally, the study uses a restricted cohort of two test subjects with four representative anatomical airspace pathways. It clearly does not capture a statistically significant range of inter-individual variability and inhalation patterns (beyond the simplified relaxed inhalation scenario); consequently, the generalizability of the current findings across wider populations is yet to be established.

\subsection{The main takeaways}

Backed by experimentally validated computational fluid dynamics simulations, this investigation emphasizes the critical role of optimizing spray device and formulation parameters to enhance targeted drug delivery within the complex anatomical landscape of the human nasal cavity. Key findings include:
\begin{itemize}
 \item   \textit{Optimal particle sizes:} Higher formulation densities increase particle inertial effects, shifting deposition loci toward anterior regions of the nasal airspace, owing to inertial impaction. When averaged across all formulation densities and airway-specific deposition trends, the particles within the size range of approximately 25 – 45 $\mu$m, combined with optimized spray angles (note the next bullet point), significantly maximize deposition at the nasopharynx (a key initial infection tissue site); see Fig.~\ref{fig9}a-b. Note also Table~\ref{table1} for the droplet size distributions in selected state-of-art spray products. The Dv$_{50}$ (a spray characteristic representing the volume median diameter of droplets in a spray plume \cite{finlay2001book}) for BiVax 200\textsuperscript{TM} and LuerVax\textsuperscript{TM} (both Aptar Pharma products; see Table~\ref{table1}) align well with our model predictions for optimal particle size range.

 \item   \textit{Plume angle optimization:} Narrower spray plume angles ($\theta$) are more effective in concentrating delivery toward the nasopharynx, reducing off-target deposition and improving therapeutic precision; see the global averaged map in Fig.~\ref{fig9}a-b. For example, with $d\in[25,45]$ $\mu$m, if $\theta \leq$~45$\degree$---then that gives a 45\% higher mean $\xi$ than its global mean (see Fig.~\ref{fig9}a-b). Such a $\theta$ (at its extremal value) would also render the conceptualized product tantalizingly close to LuerVax\textsuperscript{TM} in terms of specification (see Table~\ref{table1}). A more ambitious modification with $\theta\leq$~30$\degree$ for the same $d$ span improves $\xi$ by 76\% compared to its global mean in the parametric space explored. 

 \item   \textit{Parameter synergy:} As a specific prescription, the combination of particle sizes between 25 – 45 $\mu$m and spray plume angles $\leq$~30$\degree$ yields the highest average deposition efficiencies ($\sim$ 11.4\%).

\end{itemize}
In conclusion, this \textit{in silico} physiology-guided computational study provides a rational, simulation-informed design of spray-based intranasal drug delivery systems---to achieve maximal targeted deposition of pharmaceutics in the nasopharynx (a key infection launch site for several respiratory pathogens). Future work should focus on: incorporating tissue compliance effects, expanding to diverse anatomical variants, and conducting comprehensive toxicological assessments and safety checks for the optimized formulations and devices; the latter is especially critical in view of the elevated tissue deposition expected from the augmented spray designs.\\


\section*{Data availability statement}
All essential information is contained in the article. Supplementary information (including anatomical geometries, simulation files, postprocessing spreadsheets, and programming codes) are available via the open access repository \href{https://figshare.com/articles/dataset/Supplementary_data_for_Mechanics-guided_parametric_modeling_of_intranasal_spray_devices_and_formulations_for_targeted_drug_delivery_to_the_nasopharynx_/29640497}{figshare}, with doi: 10.6084/m9.figshare.29640497. The reader may also contact the corresponding author for any relevant data.

\section*{Ethics statement}
The study has used de-identified medical imaging data resourced from the Medical College of Wisconsin. The retrospective computational use of the anonymized and existing scans was approved with exempt status by the Institutional Review Board of South Dakota State University (vide IRB-2206003-EXM).

\section*{Author contributions}
MTH: geometry preparation, simulations, data post-processing, writing; 
AM: study design; preliminary simulations; 
MY: 3D printing, experiments, data analysis, writing; 
WOC: experiments, writing; 
MMHA: data analysis, proof-check; 
AB: preliminary simulations; 
DS: conceptualization, writing; 
GW: study design; 
JR: study design, funding acquisition;
GF: spray characterization; 
SJ: study design, funding acquisition;
JS: study design, writing; 
SB: conceptualization, study design, funding acquisition, project administration, digital reconstructions, data analysis, writing.
All authors have reviewed the manuscript.

\section*{Funding}
This work is supported by a sponsored grant from Aptar Pharma, at South Dakota State University (with a subaward at Cornell University). Supplemental support came from SB's National Science Foundation CAREER Award (\href{https://www.nsf.gov/awardsearch/showAward?AWD_ID=2339001&HistoricalAwards=false}{CBET 2339001}, from the Fluid Dynamics program, on the multiscale respiratory physics in the human upper airway). The opinions, interpretations, findings, and conclusions presented herein are solely those of the authors and do not necessarily represent the views of the relevant funders.

\section*{Acknowledgments}
SB thanks Dr.~Guilherme Garcia (Medical College of Wisconsin) for formally sharing existing, de-identified airway imaging.

\section*{Conflict of interest}
JS, GW, GF are employed by Aptar Pharma. The other authors affirm that the research was carried out without any commercial or financial relationships that could be interpreted as potential conflicts of interest.\\


\hrule

\bibliographystyle{plain}
\bibliography{mybib}

\vspace{0.5cm}


\end{document}